\documentclass[pre,twocolumn,showpacs,superscriptaddress]{revtex4}
\usepackage{amsmath}
\usepackage{bm}
\usepackage{graphicx} 
\usepackage{color}

\begin{document}

\title{The nonequilibrium glassy dynamics of self-propelled particles} 

\author{Elijah Flenner}
\affiliation{Department of Chemistry, Colorado 
State University, Fort Collins, CO 80523, USA}

\author{Grzegorz Szamel}
\affiliation{Department of Chemistry, 
Colorado State University, Fort Collins, CO 80523, USA}

\affiliation{Laboratoire Charles Coulomb, UMR 5221 CNRS, 
Universit{\'e} Montpellier, Montpellier, France}

\author{Ludovic Berthier}
\affiliation{Laboratoire Charles Coulomb, UMR 5221 CNRS, 
Universit{\'e} Montpellier, Montpellier, France}

\date{\today}


\begin{abstract}
We study the glassy dynamics taking place in dense assemblies 
of athermal active particles that are driven solely 
by a nonequilibrium self-propulsion
mechanism. Active forces are  
modeled as an Ornstein-Uhlenbeck stochastic process, 
characterized by a persistence 
time and an effective temperature, and particles 
interact via a Lennard-Jones potential that yields 
well-studied glassy behavior in the 
Brownian limit, obtained as the persistence time vanishes. 
By increasing the persistence time, the system
departs more strongly from thermal equilibrium and we provide 
a comprehensive numerical analysis of the structure and dynamics  
of the resulting active fluid. 
Finite persistence times profoundly affect 
the static structure of the fluid and give rise to nonequilibrium 
velocity correlations that are absent in thermal systems.
Despite these nonequilibrium features, for any value of the persistence time 
we observe a nonequilibrium glass transition 
as the effective temperature is decreased. Surprisingly, increasing departure from thermal
equilibrium is found to promote (rather than suppress) the glassy dynamics.
Overall, our results suggest that with increasing persistence time, 
microscopic properties of the active fluid change 
quantitatively, but the broad features of the nonequilibrium 
glassy dynamics observed with decreasing the effective temperature remain 
qualitatively similar to those of thermal glass-formers.
\end{abstract} 

\maketitle

\section{Introduction}

Systems composed of self-propelled (active) particles
represent a class of out-of-equilibrium fluids
that exhibit behavior not observed in 
other fluid systems. One well-studied example is the 
phase separation of moderately dense self-propelled fluid systems
without attractive interactions 
\cite{Fily2012,Redner2013a,Redner2013,Cates2013,Fily2014,Wysocki2014}.
It is of fundamental interest to understand differences between these
novel types of active fluids and equilibrium fluids. 
In particular, many groups strive to determine 
whether and what concepts known from the theory of equilibrium 
fluids can be used to describe   
active fluids \cite{Tailleur2008,Speck2014,Takatori2014,Wittkowski2014,Solon2015,
Farage2015,Bialke2015,Fodor2016}. One of the 
questions discussed recently is the glassy dynamics and the glass transition in self-propelled 
fluids and their comparison to the corresponding phenomena in 
equilibrium fluids  
\cite{Angelini2011,Schotz2013,Garcia2015,alberto,Gravish2015,
Henkes2011,Berthier2013,Ni2013,Berthier2014,Levis2015,Szamel2015,Szamel2016,chinese,Mandal2014,
Eaves2014,Reichhardt2014,Manning2015,farage,nandi}. 
 
Part of the motivation to analyze the glassy dynamics of 
active materials stems from experiments. 
There is ample experimental evidence that active fluids may exhibit glassy 
dynamics. Angelini \textit{et al.}\ \cite{Angelini2011} found, 
for instance, that migrating cells 
exhibited glassy dynamics, such as a diminishing self diffusion coefficient and 
heterogeneous dynamics, as the cell 
density increases. Sch\"otz \textit{et al.}\ \cite{Schotz2013} found signatures
of glassy dynamics in embryonic tissues. More recently, 
Garcia \textit{et al.}\ \cite{Garcia2015}
found an amorphous solidification process in collective motion of a 
cellular monolayer. At larger scale, ant traffic \cite{Gravish2015}
and ant aggregates \cite{alberto} were shown to display 
physical processes reminiscent of glassy solids.
To understand the glassy dynamics of cells or ant colonies, it
is important to understand what are the general features 
of glassy dynamics that are preserved when a fluid is driven 
out of equilibrium, and what new behavior may emerge that
does not have an equilibrium counterpart. 

On the theoretical front, these experimental investigations are 
complemented by a large number of simulational studies of glassy dynamics which 
have mainly focused 
on the narrower class of self-propelled particles 
\cite{Henkes2011,Berthier2013,Ni2013,Berthier2014,Levis2015,Szamel2015,Szamel2016}. 
Henkes {\it et al.} \cite{Henkes2011} analyzed how 
activity and aligning velocity interactions 
excite vibrational modes in an amorphous solid near jamming.  
Hard particles with self-propulsion were studied in 
two and three dimensions, the main outcome being that the 
equilibrium glass transition is shifted to larger densities
when either the magnitude \cite{Ni2013}
or the persistence time \cite{Berthier2014,Levis2015} 
of the self-propulsion is increased. Simulational studies of systems 
with non-aligning continuous interactions started with
Wysocki \textit{et al.}\ \cite{Wysocki2014}, who simulated a dense system of
active particles. 
Fily \textit{et al.}\ \cite{Fily2014} investigated the phase 
diagram of a dense active system as a function of density, activity
and thermal noise, and identified a glassy phase in the high-density,
small self-propulsion speed regime, but the detailed 
structure and glassy dynamics were not analyzed. 
More recently, Mandal \textit{et al.}\ \cite{Mandal2014} 
introduced an active version of the binary Lennard-Jones mixture 
(the Kob-Andersen model \cite{Kob1994}) that we also study in the present work.
They found that, upon increasing the magnitude of the self-propulsion,
the long-time dynamics speeds up leading to a disappearance of the 
glass phase, in apparent agreement with hard sphere studies. 
We shall present below a very different qualitative 
picture of the glass transition in systems of self-propelled particles.

On the theoretical side, 
Berthier and Kurchan \cite{Berthier2013} have 
used a general spin-glass model to argue that systems which 
have a glass transition in equilibrium could also exhibit a glass
transition when driven away from equilibrium through self-propulsion. 
They concluded that although the specific features of the
transition may change with the nature of the non-thermal 
driving force, such as its location, general signatures 
exhibited by thermal glass-forming systems should still be relevant.
Farage and Brader \cite{farage}
attempted to extend the mode-coupling theory
for the glass transition to account for activity, and also
concluded that self-propulsion could shift, but not destroy,
the glass transition. Very recently, Nandi \cite{nandi} proposed 
a phenomenological extension of random-first order transition 
theory to study active glasses. 

In an effort to construct a minimal non-trivial model 
to understand the competition between activity and glassy dynamics, 
we \cite{Szamel2015} have started a simulational investigation 
of a self-propelled version of the Kob-Andersen Lennard-Jones binary mixture
that is different from the model proposed in Ref.~\cite{Mandal2014}
in many aspects. Our goal is to construct a model where self-propulsion
is the unique source of motion, which neglects aligning interactions, 
obeys continuous equations of motions and where the equilibrium 
limit can be taken continuously. To this end, we consider a system
with continuous interactions in which the self-propulsion evolves 
according to an Ornstein-Uhlenbeck stochastic process.
This self-propulsion model was introduced by one of us 
earlier \cite{Szamel2014}, as a continuous 
representation of the Monte-Carlo dynamics proposed in \cite{Berthier2013}. An
essentially identical model was independently introduced and
studied by Maggi \textit{et al.}\ \cite{Maggi2015}. The general class 
of model active systems in which the self-propulsion evolves according to 
an Ornstein-Uhlenbeck process was recently named 
Active Ornstein-Uhlenbeck Particles (AOUPs) systems \cite{Fodor2016}. 
In this class of systems, the evolution of the self-propulsion is controlled by two 
parameters. An effective temperature quantifies the strength
of the self-propulsion force, and a persistence time controls 
the duration of the persistent self-propelled motion. 
One attractive feature of this model is that the departure from thermal 
equilibrium is characterized by one parameter, 
the persistence time, because in the limit of 
vanishing persistence time the system becomes equivalent to 
an equilibrium thermal system at the temperature equal to the 
effective temperature. Such continuous approach to the equilibrium 
situation is an attractive feature of the model, which is impossible 
for both active Brownian \cite{tenHagen} 
or run-and-rumble \cite{runandtumble} particle models.
This is also very different from alternative approaches in which the 
active force acts in addition to thermal motion, where fluidisation 
of the system by addition of active forces is then automatically
guaranteed \cite{farage,Mandal2014,nandi}. 

In a previous 
short account of our results \cite{Szamel2015}, we showed that with 
increasing departure from equilibrium for a range of effective
temperatures the dynamics can both speed up or slow down. We showed 
that this effect
can be qualitatively rationalized within a mode-coupling-like theory 
for the dynamics
of active systems (see Ref. \cite{Szamel2016} for a detailed presentation 
of the theory).
We found that the important ingredient of this theory was incorporation 
of the spatial 
correlations of the velocities of the individual active particles. 
These velocity
correlations have no equilibrium counterpart. Very recently, 
the existence and properties
of velocity correlations in active particles systems 
were independently studied by Marconi \textit{et al.} \cite{Marconi2016}.

In the present article, we present a very comprehensive set of numerical 
results regarding the structure and glassy dynamics 
of the self-propelled Kob-Andersen Lennard-Jones binary mixture which we introduced
earlier \cite{Szamel2015}. We focus on the dependence of the glassy dynamics
on the departure from thermal equilibrium, which we quantify by the value
of the persistence time of the self-propulsion. 
By increasing the persistence time
from zero to a finite value we continuously move from an overdamped 
thermal Brownian system through a moderately non-equilibrium system and then
to a strongly non-equilibrium self-propelled system. 
Along the way, we systematically study changes in the structural and dynamic
properties of the system. Overall, our results suggest that 
activity induces profound changes in the detailed structure 
of the nonequilibrium fluid, but the glassy dynamics 
of active particles, despite taking place far from thermal 
equilibrium \cite{Berthier2013,Levis2015},  
is qualitatively similar to that observed in equilibrium
fluids. We find that active particles display slow dynamics, 
complex time dependencies of relaxation functions, and 
spatially heterogeneous dynamics with only quantitative differences 
between equilibrium and active systems.  

The paper is organized as follows. 
In Sec.~\ref{sim} we describe the self-propulsion model and the interactions. 
In Sec.~\ref{struc} we describe how the activity 
influences the structure and find that the structure is sensitive to 
the persistence time, but weakly dependent on the effective temperature
at fixed persistence time. In Sec.~\ref{dynamics} we classify the slowing down 
at a fixed persistence time and characterize the glass transition as
a function of the effective temperatures. In Sec.~\ref{hetero}
we examine dynamic heterogeneity. We finish with a summary and the 
conclusions that can be drawn from this work.    

\section{Athermal model for self-propelled particles}

\label{sim}

We simulate a system of 
interacting self-propelled particles moving in a viscous 
medium. The motion of the particles is
solely due to an internal driving force that
evolves according to the Ornstein-Uhlenbeck process.
The absence of thermal noise qualifies our model as being `athermal'. Such model
should be applicable for large enough active particles where 
thermal noise is negligible compared to their self-propulsion.
The equations of motion for the active particles are
\begin{equation}\label{drdt}
\dot{\mathbf{r}}_i = \xi_0^{-1}\left[ \mathbf{F}_i + \mathbf{f}_i \right],
\end{equation}
where $\mathbf{r}_i$ is the 
position of particle $i$, the force 
$\mathbf{F}_i = -\sum_{j \ne i} \nabla_i V(r_{ij})$
originates from the interactions, $\mathbf{f}_i$ is the self-propulsion force,
and $\xi_0$ is the friction coefficient of an isolated particle.
Note that by using the
single-particle friction coefficient in Eq. (\ref{drdt}) we neglect 
hydrodynamic interactions. 
The equations of motion for self-propulsion force $\mathbf{f}_i$ are 
given by
\begin{equation}\label{dfdt}
\tau_p \dot{\mathbf{f}}_i = -\mathbf{f}_i + \boldsymbol{\eta}_i,
\end{equation}
where $\tau_p$ is the persistence time of the self-propulsion and 
$\boldsymbol{\eta}_i$ is a Gaussian white noise with zero mean and
variance $\left<\boldsymbol{\eta}_i(t) \boldsymbol{\eta}_j(t')
\right>_{\text{noise}} = 
2 \xi_0 T_{\mathrm{eff}} \boldsymbol{I}\delta_{ij}\delta(t-t')$, where 
$\left< ... \right>_{\text{noise}}$ 
denotes averaging over the noise distribution, $T_{\mathrm{eff}}$ is a 
single particle effective temperature,
and $\boldsymbol{I}$ is the unit tensor. 
We alert the reader that we changed the 
equation of motion for the self-propulsion, Eq. (\ref{dfdt}),
to simplify notation with respect to our earlier work \cite{Szamel2015}.
This also makes it more consistent with the 
corresponding equation of motion used by 
Fodor \textit{et al.} \cite{Fodor2016}.

The mean-squared displacement of an isolated particle
is given by
\begin{equation}
\left< \delta r^2(t) \right> = 
\frac{6 T_{\mathrm{eff}}}{\xi_0} \left[ \tau_p \left( e^{-t/\tau_p} 
- 1\right) + t \right].
\end{equation}
For $t \ll \tau_p$ the motion is ballistic with 
$\left< \delta r^2(t) \right> \approx 3 T_{\mathrm{eff}} \tau_p^{-1} t^2/\xi_0$, 
and for long times the motion is diffusive with 
$\left< \delta r^2(t) \right> \approx 6 T_{\mathrm{eff}} t/\xi_0$. 

For a many-particle system, we choose as three independent control parameters, 
the number density $\rho$, the effective temperature $T_{\mathrm{eff}}$ and 
the persistence time $\tau_p$. For a constant effective temperature,
in the limit of vanishing persistence time, the dynamics defined by 
Eqs.~(\ref{drdt}, \ref{dfdt})  
approaches overdamped Brownian dynamics 
at temperature $T = T_{\mathrm{eff}}$. Therefore, $\tau_p$ 
can be considered as the appropriate 
measure of the deviation from an equilibrium 
system undergoing overdamped Brownian motion. For this reason, we 
compare some results to simulations of the corresponding 
overdamped Brownian system \cite{Flenner2005a,Flenner2005b}. 
We note that an approximate mapping has recently been 
proposed \cite{Farage2015} between our model and the standard 
active Brownian particles model \cite{tenHagen}.

We simulate the classic Kob-Andersen 80:20 binary mixture at a 
number density $\rho = 1.2$ to study the glassy dynamics. 
At such large density, the physics of phase separation is 
irrelevant and we can focus exclusively on the glassy dynamics
of an homogeneous active fluid.
The particles interact via a Lennard-Jones potential
\begin{equation}
V_{\alpha \beta}(r_{nm}) = 4 \epsilon_{\alpha \beta} \left[ 
\left( \frac{\sigma_{\alpha \beta}}{r_{nm}} \right)^{12} -  
\left( \frac{\sigma_{\alpha \beta}}{r_{nm}} \right)^{6} \right],
\end{equation} 
where $r_{nm} = | \mathbf{r}_n - \mathbf{r}_m |$. The interaction parameters 
are $\epsilon_{BB} = 0.5 \epsilon_{AA}$, $\epsilon_{AB} = 1.5 \epsilon_{AA}$, 
$\sigma_{BB} = 0.88 \sigma_{AA}$, and $\sigma_{AB} = 0.8 \sigma_{AA}$. The 
results will be presented in reduced units where $\epsilon_{AA}$ is the unit of 
energy, $\sigma_{AA}$ is the unit of length, and 
$\sigma_{AA} \xi_0/\epsilon_{AA}$ 
is the unit of time. 
Most simulations used systems of 1000 particles with periodic 
boundary conditions. 
A few runs which focused on the small wave-vector behavior of 
certain correlation functions
used systems of 27000 particles. 
Since the equations of motion allow 
for a drift of the center of mass, we corrected for this center of 
mass drift in calculating
time correlation functions. 

We simulated persistence times $\tau_p = 2\times10^{-4}$, $5\times10^{-4}$, 
$2\times10^{-3}$, $1\times10^{-2}$, $2\times10^{-2}$, $3\times10^{-2}$, 
$5\times10^{-2}$, 
and $1\times10^{-1}$. 
For each persistence time we simulated systems for $T_{\mathrm{eff}} = 2.2$,
2.0, 1.8, 1.6, 1.4, 1.2, 1.0, 0.95, 0.9, 0.85, 0.8, 0.75, 0.7, and 0.65. For 
$\tau_p = 2\times10^{-4}$, $5\times10^{-4}$, $1\times10^{-2}$,
and $2\times10^{-2}$ we 
also simulated $T_{\mathrm{eff}} = 0.6$, 0.55, 0.5, and 0.47. For 
$\tau_p = 3\times10^{-2}$ we also simulated $T_{\mathrm{eff}} = 0.6$, 
0.55, and 0.5. 
For $\tau_p = 5\times10^{-2}$ and $\tau_p = 1\times10^{-1}$ we also 
simulated $T_{\mathrm{eff}} = 1.1$.   
This range of temperatures allowed us to study the high temperature liquid-like 
behavior of the active fluid down to the glassy dynamics regime. 

\section{Structure: equal-time positions and 
velocity correlations}
\label{struc}

Glassy dynamics is associated with a dramatic increase 
of the relaxation time with a small change of the control parameter 
(generally temperature or density) without a 
dramatic change in the structure. In this section we discuss the 
changes in the steady-state structure by 
examining two equal-time two-body distribution functions, the well-known pair 
distribution function and a novel function characterizing 
spatial correlations of
the velocities of the self-propelled particles. 

\subsection{Pair correlation functions}

\begin{figure}
\includegraphics[width=3.2in]{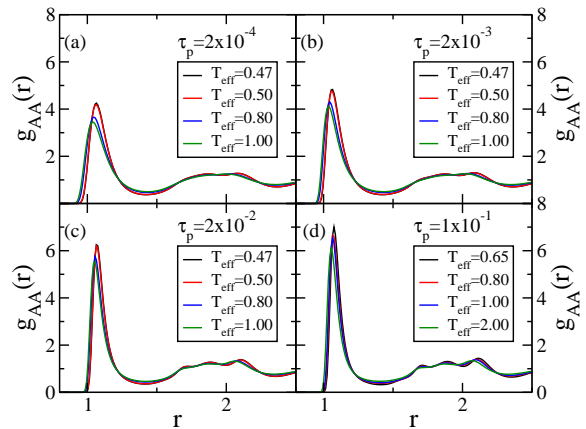}
\caption{\label{struc1}The pair distribution function $g_{AA}(r)$ calculated for the larger,
more abundant $A$ particles for (a) $\tau_p = 2\times 10^{-4}$, (b) 
$2\times 10^{-3}$, 
(c) $2\times 10^{-2}$ and (d) $1\times 10^{-1}$. For panel
panel (d), we could not equilibrate the system below $T_{\mathrm{eff}} = 0.65$, 
thus we do not show data for $T_{\mathrm{eff}} = 0.47$.
For all $\tau_p$, the weak temperature dependence contrasts with the 
profound influence of the persistence time on the pair structure.}
\end{figure}

The pair distribution function is defined as \cite{simple}
\begin{equation}
g_{\alpha \beta}(r) = \frac{V}{ N_\alpha N_\beta } \left< 
\sum_{n}^{N_\alpha} \sum_{m \ne n}^{N_\beta} 
\delta( \mathbf{r} - (\mathbf{r}_n - \mathbf{r}_m)) \right>.
\end{equation}
Shown in Fig.~\ref{struc1} is the pair distribution 
function for the larger particles, $g_{AA}(r)$, for (a) $\tau_p = 2\times10^{-4}$, (b) 
$2\times10^{-3}$, (c) $2\times10^{-2}$ and (d) $1\times10^{-1}$ for some representative 
effective temperatures. We note that $g_{AA}(r)$ for $\tau_p = 2\times10^{-4}$ is nearly 
identical to $g_{AA}(r)$ obtained over-damped Brownian dynamics 
simulations \cite{Flenner2005a,Flenner2005b}. However, for the larger persistence times there is 
an enhancement of the structure as shown by the increased height of the first peak of $g_{AA}(r)$. 
Furthermore, for the largest persistence time shown, the second 
peak begins to split into three peaks. While it is often observed that the second peak of 
$g_{AA}(r)$ begins to split into two peaks for the thermal Kob-Andersen system at low 
temperatures, the splitting into three peaks is typically not observed. 
These results suggest that the self-propulsion can promote 
the presence of structures 
that are not favored in the thermal system, 
indicating that the local structure of the active fluid 
is presumably very hard to predict from the sole knowledge of the 
pair potential. 
Importantly, the structure does not significantly change with 
decreasing $T_{\mathrm{eff}}$.
In particular, we see no evidence of 
crystallization or phase separation 
between the two species for any of the simulations. 

\begin{figure}
\includegraphics[width=3.2in]{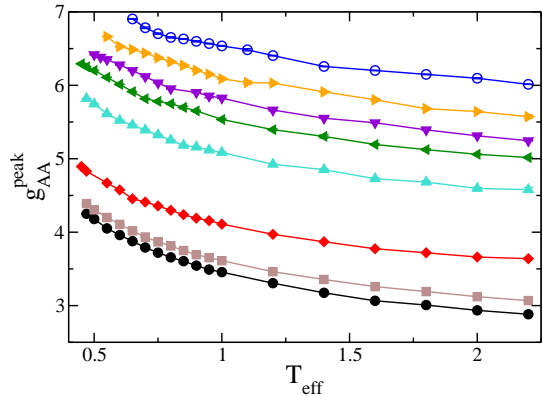}
\caption{\label{grpeak}The height of the first peak of $g_{AA}(r)$ as 
a function of the effective temperature $T_{\mathrm{eff}}$ for 
$\tau_p = 2\times10^{-4}$,
$5\times10^{-4}$, $2\times10^{-3}$, $1\times10^{-2}$, 
$2\times10^{-2}$, $3\times10^{-2}$, 
$5\times10^{-2}$, and $1\times10^{-1}$ listed from bottom to top.}
\end{figure}

To quantify the change in structure we calculated the height of the first peak of 
$g_{AA}(r)$, $g_{AA}^{\mathrm{peak}}$, as a function of $T_{\mathrm{eff}}$ for each 
persistence time. The results are shown in Fig.~\ref{grpeak}. 
For a fixed persistence time, $g_{AA}^{\mathrm{peak}}$
increases as $T_{\mathrm{eff}}$ decreases. 
At fixed $T_{\mathrm{eff}}$, $g_{AA}^{\mathrm{peak}}$
increases with increasing persistence time, which demonstrates that 
the structure is enhanced with increasing persistence times.
We note that the influence of the persistence time on the peak 
is much more pronounced that the influence of the effective temperature. 
This means that, for a given persistence time, the local structure 
changes weakly as the nonequilibrium glass transition 
temperature is approached, which is reminiscent of the equilibrium 
glass phenomenology. We note, for later reference, 
that the relative change of the peak height as temperature decreases 
is actually weaker for the largest persistence time than it is for the 
equilibrium model. 
  
\begin{figure}
\includegraphics[width=3.2in]{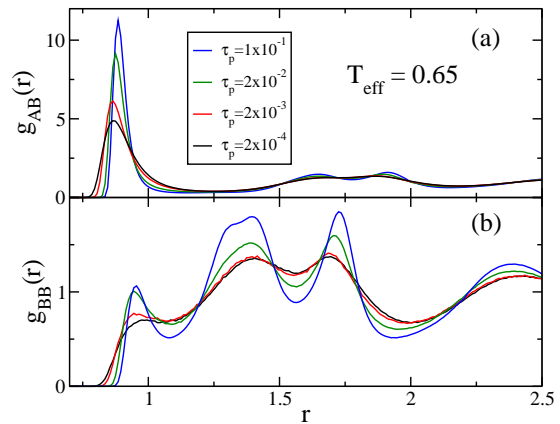}
\caption{\label{gr065}The pair distribution functions 
$g_{AB}(r)$ and $g_{BB}(r)$ 
at $T_{\mathrm{eff}} = 0.65$ for $\tau_p = 2\times10^{-4}$, 
$2\times10^{-3}$, $2\times10^{-2}$,
and $1\times10^{-1}$. Increasing the persistence time modifies 
strongly the local structure of the fluid.}
\end{figure}

Finally we also examined distribution functions involving the minority component, 
$g_{AB}(r)$ and $g_{BB}(r)$, at a fixed $T_{\mathrm{eff}} = 0.65$ 
for $\tau_p = 2\times10^{-4}$, $2\times10^{-3}$, $2\times10^{-2}$ 
and $1\times10^{-1}$. These functions are shown in Fig.~\ref{gr065}. 
Note that $T_{\mathrm{eff}} = 0.65$ is the lowest effective 
temperature where we have 
equilibrated simulations for $\tau_p = 1\times10^{-1}$. 

We find that with increasing persistence time the height of the first peak of $g_{AB}(r)$ in Fig.~\ref{gr065}(a)
increases and the splitting of the second peak becomes more pronounced.  
Furthermore, the width of the first peak decreases, which suggests an increased ordering. 
The enhanced structure is more pronounced when one examines $g_{BB}(r)$, which is shown in 
Fig.~\ref{gr065}(b). 
For $g_{BB}(r)$ there is a pronounced increase of the first four peaks. 
Notably, the first peak is generally a 
non-distinct shoulder for this model in the Brownian limit. It becomes 
much more 
pronounced with the increasing $\tau_p$. Recall that $\sigma_{BB} = 0.88$, 
thus this peak 
corresponds to an increased probability of finding another $B$ particle 
in the first neighbor 
shell of a $B$ particle. 

An interesting question, which is beyond the scope of this work, is 
how the self-propulsion influences the relative populations of 
locally favored structures \cite{paddyreview} 
of the system. Answering this question could lead to a new 
route to make desired structures more favorable in dense liquids. 
What we do observe here is 
that increasing persistence times of the self-propulsion changes 
the local structure of a dense binary fluid in a highly non-trivial manner.
This implies that the active Lennard-Jones fluid driven by nonequilibrium 
self-propulsion forces behaves a dense fluid with thermodynamic 
properties that are quantitatively distinct from the original 
equilibrium model, and it becomes a `different' fluid. 

\subsection{Velocity correlations}

Most theories of the equilibrium fluid dynamics use as their input the pair distribution
function or quantities that can be calculated from this function 
\cite{simple,Berthier2010}.
While deriving a theory for the dynamics of active systems \cite{Szamel2015,Szamel2016}, 
we discovered that the active
fluid dynamics is also influenced by correlations of velocities of individual active
particles. This influence appears in two places. First, the short-time dynamics can be analyzed
exactly and one can show that it is determined by a pair correlation function of the 
particles' velocities. Second, within an approximate theory, which is similar in spirit to
the well-known theory for the dynamics of glass forming fluids, the mode-coupling theory 
\cite{Goetzebook}, the long-time dynamics is influenced by that same correlation function. 
We should emphasize that this correlation function does not have an equilibrium analogue.
It becomes structure-less in the limit of vanishing persistence time, \textit{i.e.}
when our system becomes equivalent to an equilibrium (thermal) system. We should also 
mention that the existence and importance of velocity correlations in active particles systems 
were also stressed recently by Marconi \textit{et al.} \cite{Marconi2016}.

Our theoretical analysis therefore leads 
us to the following correlation function of the particle velocities,
\begin{equation}\label{velcor}
\omega_{\parallel}(q) = \frac{1}{N \xi_0^{2}} 
\left<\left|\hat{\mathbf{q}} \cdot \sum_i \left(\mathbf{F}_i+\mathbf{f}_i\right) 
e^{-i \mathbf{q}\cdot\mathbf{r}_i}\right|^2\right>.
\end{equation}
In Eq.~(\ref{velcor}), $\xi_0^{-1}\left(\mathbf{F}_i+\mathbf{f}_i\right)$
is the velocity of particle $i$, see Eq. (\ref{drdt}), 
and thus $\omega_{\parallel}(q)$
quantifies longitudinal
spatial correlations of the velocities of the individual particles. 

It turns out that both the wave-vector dependence of $\omega_{\parallel}(q)$ and its
overall magnitude change with our two control parameters, $T_{\mathrm{eff}}$ and
$\tau_p$. For this reason, we first show the changes in the wave-vector dependence
by presenting velocity correlations normalized by their infinite wave-vector limit, 
$\omega_{\parallel}(q)/\omega_{\parallel}(\infty)$, and then we show the 
changes in the overall magnitude by presenting the infinite wave-vector limit, 
$\omega_{\parallel}(\infty)$. We note that 
$\omega_{\parallel}(\infty)$ is related to the average magnitude of the velocity,
\begin{equation}
\omega_{\parallel}(\infty) = \frac{1}{3 N \xi_0^{2}}
\left<\sum_i \left(\mathbf{F}_i+\mathbf{f}_i\right)^2 \right>. 
\end{equation}

\begin{figure}
\includegraphics[width=3.2in]{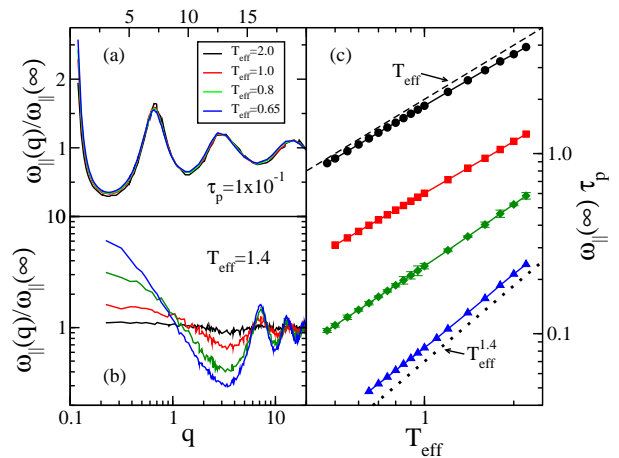}
\caption{\label{omega} (a) Velocity correlations as 
measured by $\omega_\parallel(q)/\omega_\parallel(\infty)$ for $T_{\mathrm{eff}} = 
2.0$, 1.0, 0.8, and 0.65
for $\tau_p = 1\times10^{-1}$. (b) Velocity correlations as measured by 
$\omega_\parallel(q)/\omega_\parallel(\infty)$
for $\tau_p = 2\times10^{-4}$, $2\times10^{-3}$, $2\times10^{-2}$, and $1\times10^{-1}$ at a fixed 
$T_{\mathrm{eff}} = 1.4$. 
(c) $\omega_{\parallel}(\infty) \tau_p$ for $\tau_p = 2\times10^{-4}$, 
$2\times10^{-3}$, $2\times10^{-2}$, and $1\times10^{-1}$ as a function of $T_{\mathrm{eff}}$.}
\end{figure}

In Fig.~\ref{omega}(a) we show $\omega_{\parallel}(q)/\omega_{\parallel}(\infty)$ for 
$\tau_p = 1\times10^{-1}$ at $T_{\mathrm{eff}} = 2.0$, 1.0 , 0.8, and 0.65. 
We find that, possibly except at the smallest wave-vectors, there is very little change 
in $\omega_{\parallel}(q)/\omega_{\parallel}(\infty)$ with decreasing $T_{\mathrm{eff}}$ at 
fixed $\tau_p$, and we find this to be the case for every $\tau_p$. However, as shown in 
Fig.~\ref{omega}(b), there is a large change in $\omega_{\parallel}(q)/\omega_{\parallel}(\infty)$ 
with $\tau_p$ at constant $T_{\mathrm{eff}} = 1.4$. 
Thus, the increase of the velocity correlations is a consequence of the increased 
persistence time, but this structural measure also does not change significantly with decreasing 
effective temperature.  This trend is similar to the 
observation made for the pair distribution function.

We note two different aspects of the increasing velocity correlations. First, 
$\omega_{\parallel}(q)/\omega_{\parallel}(\infty)$ develops finite wave-vector oscillatory
structure which, roughly speaking, follows that of the static structure factor. Second, 
$\omega_{\parallel}(q)/\omega_{\parallel}(\infty)$ develops a peak at the zero wave-vector,
which indicates growing velocity correlation length. A detailed investigation of this latter
feature is left for a future study. Here we only note that at $T_{\mathrm{eff}}=1.4$ and 
$\tau_p = 1\times10^{-1}$, which corresponds to a liquid state, this length reaches the
value of about 2.6 particle diameter (recall that this correlation length
scale is not even defined for Brownian particles). 
While still modest, such a correlation length is longer 
than any other structural length
in an equilibrium fluid with similar dynamics. Thus, increasing 
the persistence time induces the emergence of non-trivial 
spatial correlations in the active fluid that are fully 
nonequilibrium in nature.

In Fig.~\ref{omega}(c) we present the dependence of the overall magnitude of 
$\omega_{\parallel}(q)$ on $T_{\mathrm{eff}}$ and $\tau_p$ by plotting
$\tau_p \omega_{\parallel}(\infty)$. We found \cite{Szamel2016} that this 
quantity is a measure of the intermediate time dynamics (under the assumption that there
is a separation of time scales between short-time ballistic motion and the long-time
motion heavily influenced by the time-dependent internal friction). We show 
the dependence of $\tau_p \omega_{\parallel}(\infty)$ on $T_{\mathrm{eff}}$ for 
$\tau_p = 2\times10^{-4}$, $2\times10^{-3}$, $2\times10^{-2}$, $1\times10^{-1}$ listed from top 
to bottom. Since at a given effective temperature $T_{\mathrm{eff}}$ the strength of
the self-propulsion force is $|\mathbf{f}| \sim \sqrt{T_{\mathrm{eff}}/\tau_p}$,
an increase of the $\tau_p$ at fixed $T_{\mathrm{eff}}$ will result in an overall decrease of 
$\omega_{\parallel}(\infty)$. If the self-propulsion force dominates the velocity correlations, 
one would expect that $\tau_p \omega_\parallel(\infty) \sim \mathbf{f}^2$, 
thus $\tau_p \omega_\parallel(\infty) \sim T_{\mathrm{eff}}$, 
and we find that there is an approximate linear decrease of 
$\tau_p \omega_\parallel(\infty)$ with $T_{\mathrm{eff}}$ for $\tau_p = 2\times10^{-4}$. 
However, $\tau_p \omega_\parallel(\infty)$ no longer linearly decreases linearly with decreasing 
$T_{\mathrm{eff}}$ for larger $\tau_p$, and 
$\tau_p \omega_\parallel(\infty) \sim T_{\mathrm{eff}}^{1.4}$
for $\tau_p = 1\times10^{-1}$.  
Therefore, there is an increased influence of the particle 
interactions with increasing persistence time, which is  consistent 
with findings of Marconi \textit{et al.} \cite{Marconi2016}. 
Physically, this implies that the intermediate-time dynamics 
slows down more rapidly with decreasing the effective temperature 
for large persistence times than for small ones. 
  
\section{Dynamic slowing Down and the glass transition}
\label{dynamics}

One fundamental feature of glassy dynamics
is the dramatic slowing down of the 
dynamics accompanied by little change in the two-body static structure. 
In the previous section we found little change in the structure with decreasing 
the effective temperature $T_{\mathrm{eff}}$ at fixed persistence 
time $\tau_p$. In this
section we examine the dramatic slow down of the dynamics upon changing the
control parameters in the same way, \textit{i.e.} with decreasing 
the effective temperature at fixed persistence 
time. 

\subsection{Self-intermediate scattering function}

\begin{figure}
\includegraphics[width=3.2in]{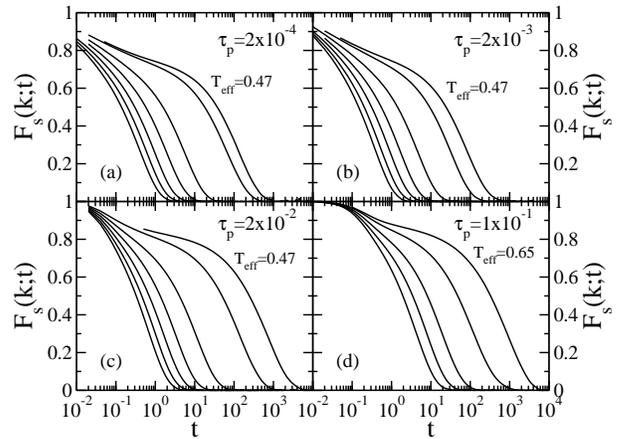}
\caption{\label{fs}The self-intermediate scattering function $F_s(k;t)$ 
from $T_{\mathrm{eff}} = 2.2$ down 
to the lowest temperature we could equilibrate for 
(a) $\tau_p = 2\times10^{-4}$, (b) $\tau_p = 2\times10^{-3}$, (c) $\tau_p = 2\times10^{-2}$,
and (d) $\tau_p = 1\times10^{-2}$
for all the simulated temperatures listed in Sec.~\ref{sim}. 
In all cases, nonequilibrium 
glassy dynamics is observed as $T_{\mathrm{eff}}$ decreases.}
\end{figure}

To examine the dynamics we study the self-intermediate scattering function
for the $A$ particles
\begin{equation}\label{selfscatt}
F_s(k;t) = \frac{1}{N_A} \left< \sum_{n=1}^{N_A} 
e^{i \mathbf{k} \cdot (\mathbf{r}_n(t) - \mathbf{r}_n(0))} \right>.
\end{equation}
While we only present results for the more abundant, larger $A$ particles, 
similar conclusions are drawn if one examines the
dynamics of the smaller $B$ particles. We choose $k = 7.2$, which is around the first peak of the 
partial static structure factor calculated for the $A$ particles. 
Shown in Fig.~\ref{fs} is $F_s(k;t)$ for representative values of the 
persistence time. 
While we observed little change in the structure with
decreasing $T_{\mathrm{eff}}$ at fixed $\tau_p$, we do see a dramatic slowing down
of the dynamics, in the sense that 
the relaxation becomes very slow at low temperatures. 
Also, as is commonly observed in equilibrium supercooled liquids,
the functional form of the time decay of the correlation functions 
is relatively independent of temperature, and is well-described by 
a stretched exponential form that also depends only weakly on the persistence
time. Therefore, we now focus on the relaxation time $\tau_\alpha$
of $F_s(k;t)$ which we define 
as when $F_s(k;\tau_\alpha) = e^{-1}$. 

\subsection{Relaxation times}

\begin{figure}
\includegraphics[width=3.2in]{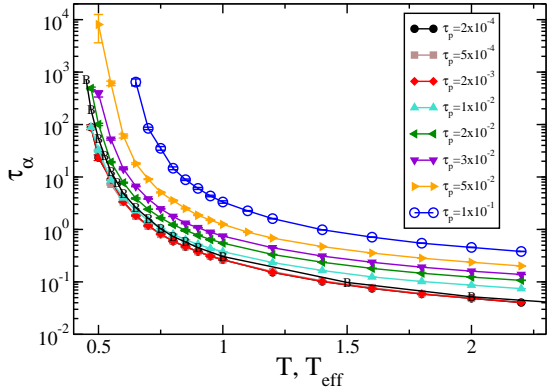}
\caption{\label{tau}The dependence of the relaxation time $\tau_\alpha$ on 
$T_{\mathrm{eff}}$ for all the persistence times studied in this work. The 
results for over-damped Brownian dynamics from Ref.~\cite{Flenner2005a} 
is shown using `B' symbols. 
Note that on the scale of the figure the results for the three shortest 
persistence times are almost identical, whereas larger 
persistence times seem to promote, rather than suppress, 
the glassy dynamics.}
\end{figure}

Shown in Fig.~\ref{tau} is the $\alpha$ relaxation time, $\tau_\alpha$, 
as a function of 
the effective temperature, $T_{\mathrm{eff}}$, for all the 
persistence times studied in this work. Also shown with 
`B' symbols is $\tau_\alpha$ for overdamped
Brownian dynamics simulations of the same 
system \cite{Flenner2005a,Flenner2005b}. 
Note that at small $\tau_p$ and a fixed $T_{\mathrm{eff}}$, 
the active system relaxes faster than the thermal 
Brownian system at $T=T_{\mathrm{eff}}$. The relaxation time becomes larger than that of 
the over-damped Brownian system for $\tau_p \ge 2\times10^{-2}$. This non-monotonic dependence of 
the relaxation time on the persistence time was reported previously, and a mode-coupling like 
theory was developed that also predicted a non-monotonic change of 
the relaxation time with 
the persistence time \cite{Szamel2016,Szamel2015}. The physics behind
the non-monotonic behavior is a competition 
between increasing structure in the velocity correlations 
(speeding up the dynamics) and an increase in the local structure (slowing
down the dynamics) as $\tau_p$ increases. 

Below we shall analyze in great detail the dramatic increase 
of $\tau_\alpha$ with decreasing effective temperature. First, it is useful to 
notice that the dynamics in the high effective temperature liquid
also bears a non-trivial dependence on the persistence time. 
For $T_{\rm eff}=2.2$, for instance, we observe that 
$\tau_\alpha$ increases by about one order of magnitude between 
Brownian dynamics and our largest simulated $\tau_p$. This change 
mirrors the increase in $\omega_\parallel(\infty)\tau_p$ 
reported in Fig.~\ref{omega}(c). This is reasonable as  
$\omega_\parallel(\infty)\tau_p$ controls the intermediate-time dynamics,
and it thus affects also the structural relaxation time in the 
high-temperature fluid, where glassy effects at long times are absent.

In contrast to the small changes in structure with decreasing effective 
temperature shown in the previous section, Sec.~\ref{struc}, there 
is a very large increase in the relaxation time with a small decrease of 
the effective temperature. We note that changing $\tau_p$ 
can dramatically change the structure, and thus it makes 
sense to examine the glass transition at a fixed $\tau_p$ and
using $T_{\mathrm{eff}}$ as the control parameter. 

Finally, we note that increasing departure from the thermal equilibrium 
does not `fluidize' the glassy system under study, 
and neither does the glass transition disappear completely, 
contrary to previous reports \cite{Mandal2014,nandi}. 
The difference stems from the fact that in the model 
of \cite{Mandal2014}, active forces act in addition to 
Brownian noise, so that the bath temperature is in fact 
not the correct control parameter to describe the glassy dynamics. 

\subsection{Mode-coupling analysis of relaxation times}

\begin{figure}
\includegraphics[width=3.2in]{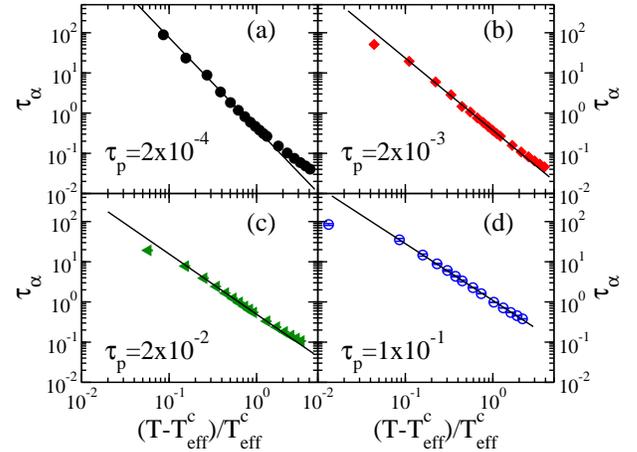}
\caption{\label{mctfit}Mode-coupling-like power law fits 
$\tau_\alpha\propto(T_{\mathrm{eff}}-T_\mathrm{eff}^c )^{-\gamma}$
for (a) $\tau_p = 2\times10^{-4}$, (b) $2\times10^{-3}$, (c) $2\times10^{-2}$, 
and (d) $1\times10^{-1}$. In all cases, there is a modest intermediate 
time regime where the power-law fit holds, as commonly found in 
equilibrium supercooled liquids.}
\end{figure}

One common way to examine glassy dynamics in thermal systems is to investigate power-law
fits of the relaxation time inspired by the mode-coupling theory \cite{Goetzebook}. These
fits result in the so-called mode-coupling transition temperature $T^c$, which
is interpreted as a crossover between a moderately supercooled regime where mode-coupling
theory is applicable and a strongly supercooled regime where the dynamics is thought to be 
dominated by somewhat vaguely defined hopping events. The development of a mode-coupling-like
dynamic theory for active systems \cite{Berthier2013,Szamel2015,Szamel2016} suggests that 
similar analysis could be performed for active glassy dynamics. 

We examine a mode-coupling like regime by fitting 
$\tau_\alpha = a(T_{\mathrm{eff}}/T_\mathrm{eff}^c - 1)^{-\gamma}$
for $T_{\mathrm{eff}}$ lower than the onset temperature, which we define 
as the highest 
temperature where $\tau_\alpha$ deviates from the high temperature Arrhenius behavior (discussed
further below). Shown in Fig.~\ref{mctfit} are the fits for four representative
persistence times. For each persistence time the mode-coupling like fits provide a reasonable 
description of the data for around two decades of slowing down. Like for thermal systems, 
there are deviations from the mode-coupling  
behavior at the lowest effective temperatures
that we could simulate. 
Notice that the quality (or poorness) of such power law fit does not 
seem to be very different between Brownian and self-propelled dynamics. 
A different result was obtained for self-propelled disks where the 
quality of the fits seemed to improve with increasing the persistence
time \cite{Berthier2014}. The present result for the binary Lennard-Jones model
then provides a counter-example suggesting that the finding for the 
hard disk system is not universal.    

\begin{figure}
\includegraphics[width=3.2in]{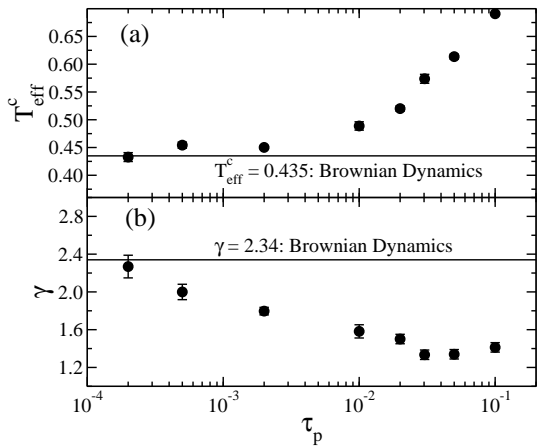}
\caption{\label{mctparam}Mode-coupling effective temperature $T_\mathrm{eff}^c$
and scaling exponent $\gamma$ obtained from power-law fits 
to $\tau_\alpha \propto (T_{\mathrm{eff}} - T_{\mathrm{eff}}^c)^{-\gamma}$
as a function of $\tau_p$. The horizontal lines are the results 
for the over-damped Brownian dynamics simulations taken 
from Ref.~\cite{Flenner2005a}. The increase of $T_{\rm eff}^c$ with $\tau_p$ 
directly indicates that self-propulsion promotes glassy dynamics, whereas
the strong change in $\gamma$ is correlated to the strong change in the 
local structure observed at large persistence times.}
\end{figure}

Shown in Fig.~\ref{mctparam}(a) are the mode-coupling effective temperatures, 
$T_{\mathrm{eff}}^c$, resulting from power-law fits as a function of the 
persistence time $\tau_p$. 
For small $\tau_p$, $T_{\mathrm{eff}}^c$ is 
close to the value of 
$T^c \approx 0.435$ found in previous studies of the Kob-Andersen 
system undergoing 
Newtonian \cite{Kob1994} or Brownian dynamics \cite{Flenner2005a}.
For larger $\tau_p$, the mode-coupling effective temperature increases monotonically 
with increasing persistence time.

Our somewhat surprising finding is thus that moving away from equilibrium 
by increasing the persistence time promotes the glassy
dynamics, which occurs at larger values of the effective
temperatures. Therefore the self-propelled Lennard-Jones system
appears more glassy than the Brownian version of the same 
interaction potential. This finding was not fully expected as
previous studies using hard potentials
have reported that the glassy dynamics is pushed to larger
densities \cite{Ni2013,Berthier2014}. The present result then suggests 
that, again, the finding for hard particles 
is not universal and the opposite 
situation is in fact possible. This was suggested theoretically on general grounds 
in \cite{Berthier2013}. Physically these findings
show that the manner in which the glass transition 
shifts when departing from thermal equilibrium stems from 
a non-trivial combination of how the local structure 
and velocity correlations are affected by the self-propulsion.
No obvious qualitative guess can be made and details of the original
Brownian system matter. 

Shown in Fig.~\ref{mctparam}(b) is the mode-coupling exponent $\gamma$ as a function of $\tau_p$. 
For $\tau_p = 2\times10^{-4}$, $\gamma$ is equal to the value obtained from Brownian 
dynamics simulations \cite{Flenner2005a}. With increasing $\tau_p$ there is 
an initial decrease in $\gamma$ until $\tau_p \approx 3\times10^{-2}$, 
then $\gamma$ is 
approximately constant for the largest persistence times studied in this work. 
In the equilibrium theory, the exponent $\gamma$ follows, in a  
non-trivial way, from the static structure of the fluid \cite{Goetzebook}. 
The situation is unclear for the nonequilibrium dynamics but the evolution 
of $\gamma$ appears qualitatively consistent with the dramatic 
change in the local structure observed in the pair correlation
function shown in Fig.~\ref{struc1}. 

\subsection{Fragility and activated dynamics}

\begin{figure}
\includegraphics[width=3.2in]{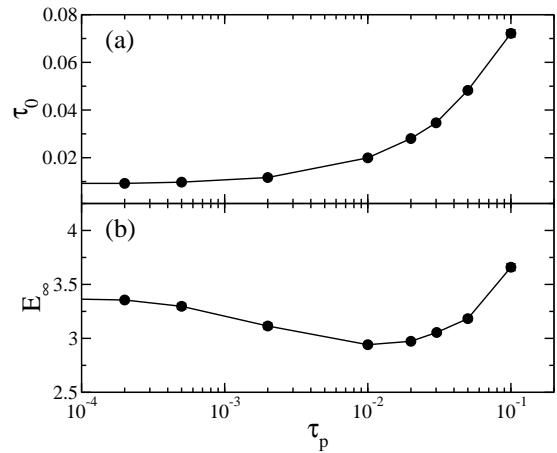}
\caption{\label{Arrparam}The time constant $\tau_0$ and the 
activation energy $E_\infty$ 
resulting from the Arrhenius-like fit
$\tau_\alpha = \tau_0 \exp(E_\infty/T_{\mathrm{eff}})$. The fits 
were performed for $T_{\mathrm{eff}} \ge 1.2$ in the high-temperature regime 
for each persistence time. }
\end{figure}

A different examination of the glassy dynamics focuses on the departures from
the high temperature Arrhenius behavior (see, \textit{e.g.} Ref. \cite{Berthier2009}). 
To this end we start by fitting the relaxation time to the Arrhenius-like formula,   
$\tau_\alpha = \tau_0\exp[E_{\infty}/T_{\mathrm{eff}}]$,
for $T_{\mathrm{eff}} \ge 1.2$ and fixed $\tau_p$ to  
obtain the  time constant $\tau_0$ and the activation energy  
$E_\infty$, see Fig.~\ref{Arrparam}, at each persistence time. We see that $\tau_0$ 
grows monotonically with increasing persistence time
in a manner which is again reminiscent of the 
changes observed in $\tau_p \omega_\parallel(\infty)$
in Fig.~\ref{omega}(c) and of the high-temperature liquid dynamics 
in Fig.~\ref{tau}. 

In contrast, the effective activation energy $E_\infty$ 
initially decreases with increasing persistence time
and then increases at higher persistence times. 
The minimum in $E_\infty$ is around $\tau_p = 1\times10^{-2}$. 
This non-monotonic dependence of $E_\infty$ is 
mirrored by the non-monotonic dependence of $\tau_\alpha$ 
versus $\tau_p$ identified in Ref. \cite{Szamel2015} and 
already discussed above. Note that $E_\infty$ characterizes the
dynamics of the high-temperature active fluid, which exhibits a non-monotonic evolution
with $\tau_p$.

\begin{figure}
\includegraphics[width=3.2in]{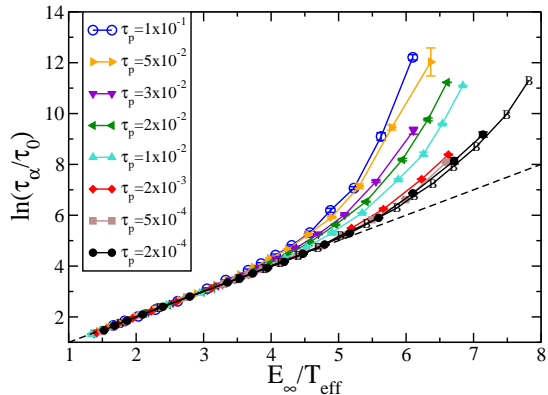}
\caption{\label{Arr} Relaxation times rescaled by the high-temperature 
Arrhenius fit $\tau_\alpha = \tau_0 \exp(E_\infty/T_{\mathrm{eff}})$ for 
all the persistence times studied in this work. A smooth monotonic 
evolution of the glassy dynamics with $\tau_p$ is observed, where
more persistent particles display steeper temperature dependence, 
i.e. they become more fragile. The Brownian results are shown 
with `B' symbols.}
\end{figure}

In Fig.~\ref{Arr}, we show that, 
once the high-temperature Arrhenius regime has been scaled out, the glassy dynamics
evolves monotonically with the persistence time. To this end, we plot    
$\ln (\tau_\alpha/\tau_0)$ versus $E_{\infty}/T_{\mathrm{eff}}$. 
When rescaled by the high temperature behavior of the 
active liquid, we now find that the slowing 
down, compared to the high temperature dynamics, monotonically 
increases with increasing 
$\tau_p$. Furthermore, the liquid dynamics appear to 
be more fragile \cite{Berthier2011}
with increasing $\tau_p$, in the sense that 
the temperature dependence in the glassy regime becomes
much steeper for larger persistence times. 

\begin{figure}[b]
\includegraphics[width=3.2in]{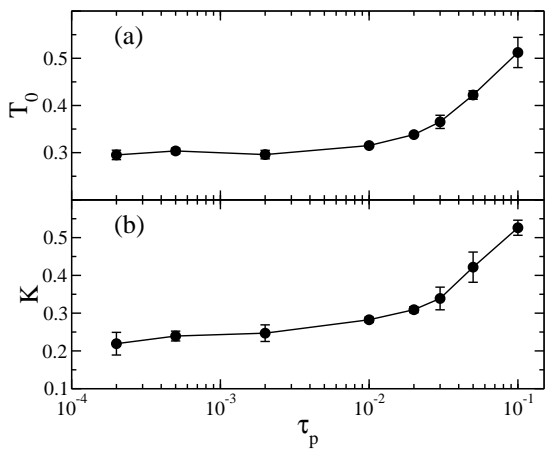}
\caption{\label{VF} (a)
The glass transition temperature $T_0$ and
(b) the fragility parameter $K$ obtained from Vogel-Fulcher-like fits
$\ln(\tau_\alpha) = a+1/[K(T_{\mathrm{eff}}/T_0 - 1)]$ as a function of 
persistence time. Both the glass transition temperature $T_0$ and the 
fragility parameter $K$ monotonically increase with increasing the 
persistence time.}
\end{figure}

To quantify the kinetic fragility and examine the glass
transition temperature, we fit the relaxation time to 
a Vogel-Fulcher-like equation,
$\ln(\tau_\alpha) = a+1/[K(T_{\mathrm{eff}}/T_0 - 1)]$,
for $T_{\mathrm{eff}} \le 0.85$.  
We find that the glass 
transition temperature $T_0$ is approximately constant with
increasing $\tau_p$ until $\tau_p \ge 2\times10^{-2}$, 
see Fig.~\ref{VF}(a). 
Note that this is about the same persistence time in which we observe a slowing
down compared to the overdamped Brownian dynamics simulation. 
For smaller $\tau_p$ the transition temperature is close
to the value known from equilibrium studies of the model. 
For larger $\tau_p$ the glass transition temperature $T_0$ increases up to 
$T_0 = 0.51 \pm 0.03$ for $\tau_p = 1\times10^{-1}$, which 
again shows that persistent motion promotes glassy dynamics 
in the present system.  

The fragility parameter 
$K$ is larger for systems that are more fragile. As suggested
by the dependence on the effective temperature of $\tau_\alpha$ in 
Fig.~\ref{Arr}, 
the fragility increases with increasing $\tau_p$, see Fig.~\ref{VF}(b).
Again, this result contrasts strongly with the results in
\cite{Mandal2014}.

A deep understanding of the kinetic fragility for equilibrium 
supercooled liquids is not available \cite{Berthier2011}, 
therefore it is difficult 
to interpret the evolution of $K$ for the present nonequilibrium 
situation. The monotonic evolution of $K$ with $\tau_p$ again confirms
the smooth evolution of glassy dynamics with the degree of 
departure from equilibrium. A large change in fragility is also 
consistent with the finding that self-propulsion dramatically 
changes the structure of the fluid, and the above conclusion 
that self-propulsion produces a `different' liquid whose 
glassy dynamics only differs in its details (such as 
kinetic fragility and glass temperature), as compared 
to typical equilibrium supercooled liquids. 

\subsection{Mean-squared displacements and 
Stokes-Einstein relation}

\begin{figure}
\includegraphics[width=3.2in]{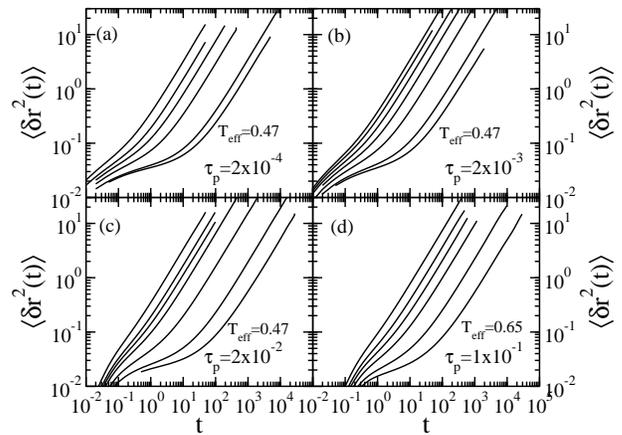}
\caption{\label{msd}The mean-squared displacement 
$\left< \delta r^2(t) \right>$ 
for (a) $\tau_p = 2\times10^{-4}$, (b) $\tau_p = 2\times10^{-3}$, (c) 
$\tau_p = 2\times10^{-2}$,
and (d) $\tau_p = 1\times10^{-1}$. 
For panels (a)-(c) $T_{\mathrm{eff}} = 0.47$, 0.5, 0.6, 0.7, 0.8, 0.9, 
and 1.0 are shown. 
In panel (d) $T_{\mathrm{eff}} = 0.65$, 0.7, 0.8, 0.9, 1.0, and 1.1 are shown.}
\end{figure}

We now examine the mean-squared displacement
$\left< \delta r^2(t) \right> = 
N^{-1} \left< \sum_i |\mathbf{r}_i(t) - \mathbf{r}_i(0)|^2 \right>$, which 
is shown in Fig.~\ref{msd} for various $\tau_p$ values. 
At short times we see a ballistic motion which results from the 
finite persistence time of the self-propulsion.
This regime is followed by a crossover to 
diffusive motion at long times, which defines the self-diffusion 
coefficient $D$ of the model. The data in Fig.~\ref{msd}
reveal that $D$ decreases dramatically as the temperature is decreased,
which is another well-known characteristic signature of the glass 
transition \cite{Berthier2011}.

Between the ballistic motion and diffusive motion a 
plateau emerges, which indicates a strongly sub-diffusive regime
whose duration increases rapidly as the temperature decreases. 
For $\tau_p = 2\times10^{-4}$
this plateau is similar to what is found for Brownian dynamics. 
For increasing $\tau_p$ the 
plateau is less pronounced and appears to occur at smaller values of 
$\left< \delta r^2(t) \right>$ for increasing $\tau_p$. 
To quantify this observation we defined the plateau 
in $\left< \delta r^2(t) \right>$ 
as the value of $\left< \delta r^2(t) \right>$ at 
the inflection point in the plot of $\ln( \left< \delta r^2(t) \right>)$ 
versus $\ln(t)$ 
at $T_{\mathrm{eff}} = 0.65$. We 
found that the plateau was around 0.035 for $\tau_p = 2\times10^{-4}$ and was 
approximately constant until $\tau_p = 1\times10^{-2}$ where it decreased 
to 0.0164 at $\tau_p = 1\times10^{-1}$. 

In equilibrium systems at temperatures above the 
onset of supercooling it is generally found that 
the Stokes-Einstein relation $D \sim \tau_\alpha^{-1}$ holds.
Therefore, $D\tau_\alpha$ is 
approximately constant for high temperature liquids. The
Stokes-Einstein relation has
been found violated in supercooled liquids, resulting in a growth 
of $D\tau_\alpha$ below the onset temperature \cite{Berthierbook,Ediger2000}. 
We examined the Stokes-Einstein
relation for our model of self-propelled particles.
Recall that as $\tau_p$ goes to zero for
a fixed effective temperature, the dynamics becomes identical to 
over-damped Brownian dynamics. 

\begin{figure}
\includegraphics[width=3.2in]{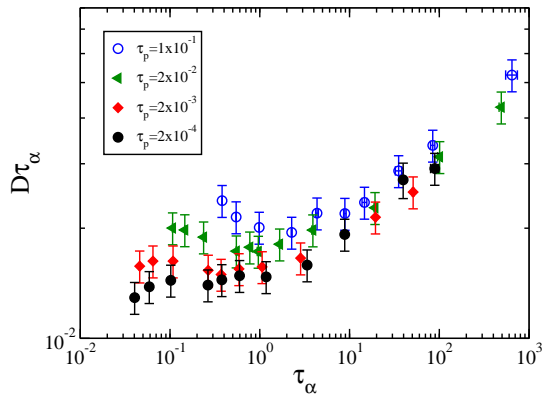}
\caption{\label{SE}Examination of the Stokes-Einstein
 relation, $D \tau_\alpha \sim const$. 
For $\tau_p = 2\times10^{-4}$ 
and $2\times10^{-3}$ $D\tau_\alpha$ is approximately constant 
for small $\tau_\alpha$, which 
corresponds to high $T_{\mathrm{eff}}$. However, for long 
persistence times $\tau_p$ the
Stokes-Einstein relation is not valid for any 
temperature range. In the glassy regime, a similar degree of Stokes-Einstein 
decoupling is observed for all $\tau_p$ values.}
\end{figure}

To examine the Stokes-Einstein relaxation 
we first calculated the diffusion coefficient using 
$D = \lim_{t \rightarrow \infty} \left< \delta r^2(t) \right>/6$. 
Shown in Fig.~\ref{SE} is the evolution of $D\tau_\alpha$ versus $\tau_\alpha$ 
for several $\tau_p$ and effective temperatures.
The main observation is that the product  $D\tau_\alpha$ changes by less 
than one decade for all systems, which indicates that the 
decrease of the diffusion and the increase of the 
relaxation time are very strongly correlated. 

We do see, however, deviations from the Stokes-Einstein 
relation. For small values of the persistence times the Stokes-Einstein relation  
is approximately valid at high effective temperatures, in other words 
for small $\tau_\alpha$. 
However, for the largest $\tau_p$ we do not find a clear 
region of effective temperatures where the
Stokes-Einstein relation is valid. Instead, we find a minimum in 
the plot of $D\tau_\alpha$ versus $\tau_\alpha$  for $\tau_p = 1\times10^{-1}$. 
At large $\tau_\alpha$, it appears that $D\tau_\alpha$ for all 
systems approximately follow the same master-curve. 
This suggests that despite small differences in the high temperature
liquid, the glassy dynamics of both Brownian and strongly self-propelled
systems shows a similar degree of Stokes-Einstein decoupling.
In particular, the (weak) correlation discussed for equilibrium 
liquids between kinetic fragility and Stokes-Einstein 
decoupling \cite{Ediger2000} is not observed here. 

\section{Dynamic Heterogeneity}
\label{hetero}

An extensively studied feature of the dynamics of supercooled liquids 
is the emergence of spatially 
heterogeneous dynamics \cite{Berthier2011,Berthierbook,Ediger2000}. 
In high temperature simple fluids, particle displacements are 
nearly Gaussian, but in supercooled liquids it has been observed that 
there are subsets of particles whose displacements are either much
greater than (fast particles) or much smaller than (slow particles) 
what would be expected from 
a Gaussian distribution of particle displacements. Furthermore, 
the fast and slow particles form transient clusters whose maximum 
characteristic 
size increases with decreasing temperature. 

\begin{figure}
\includegraphics[width=3.2in]{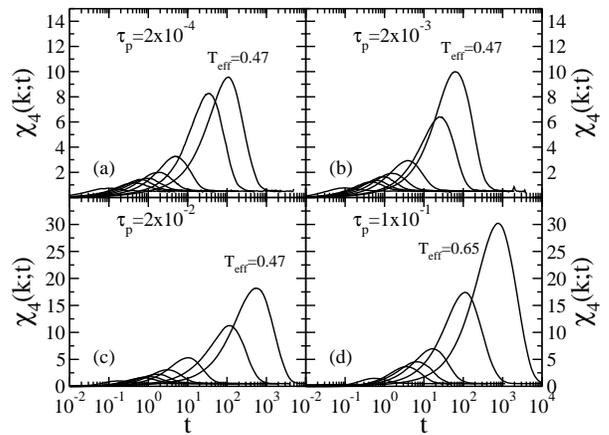}
\caption{\label{suss}The dynamic susceptibility $\chi_4(k;t)$ for 
(a) $\tau_p = 2\times10^{-4}$, (b) $2\times10^{-3}$, (c) $2\times10^{-2}$, 
and (d) $1\times10^{-1}$. 
The peak positions are around $\tau_\alpha$, and the peak height 
grows with decreasing 
$T_{\mathrm{eff}}$ for a fixed $\tau_p$. For a fixed $T_{\mathrm{eff}}$ the peak 
height grows with increasing $\tau_p$. Notice the change of scale on 
the y axis for $\tau_p = 2\times10^{-2}$ and $\tau_p = 1\times10^{-1}$. }
\end{figure}

In this section we investigate whether the self-propulsion influences 
the dynamic heterogeneity. 
To quantify such dynamic heterogeneity we monitor the fluctuations of the 
real part of the 
microscopic self-intermediate scattering function 
\begin{equation}
Re \hat{F}_s^n(k;t) = \cos\left(\mathbf{k} \cdot (\mathbf{r}_n(t) - 
\mathbf{r}_n(0))\right)
\end{equation}
of the larger $A$ particles. We define the dynamic 
susceptibility \cite{Berthierbook,Flenner2011}
\begin{eqnarray}
&& \chi_4(k;t) 
\\ \nonumber && 
= \frac{1}{N_A} \left[ \left< \left( \sum_{n=1}^{N_A} Re 
\hat{F}_s^n(k;t)\right)^2 \right> 
- \left< \sum_{n=1}^{N_A} Re \hat{F}_s^n(k;t) \right>^2 \right],
\end{eqnarray}
where we fix $k = 7.2$ as for the self-intermediate scattering function.

Shown in Fig.~\ref{suss} is the dynamic susceptibility 
$\chi_4(k;t)$ for different 
effective temperatures 
for the same persistence times as shown in Fig.~\ref{fs}. We find that the 
susceptibility grows with time, peaks at a time around $\tau_\alpha$, 
and finally decays.
With decreasing $T_{\mathrm{eff}}$ at constant $\tau_p$, the
dynamic susceptibility increases, in a way qualitatively similar 
to that in thermal systems
\cite{Flenner2011}.  This observation shows that 
the glassy dynamics observed for self-propelled particles does not 
appear very distinct from the equilibrium glassy dynamics. 

\begin{figure}
\includegraphics[width=3.2in]{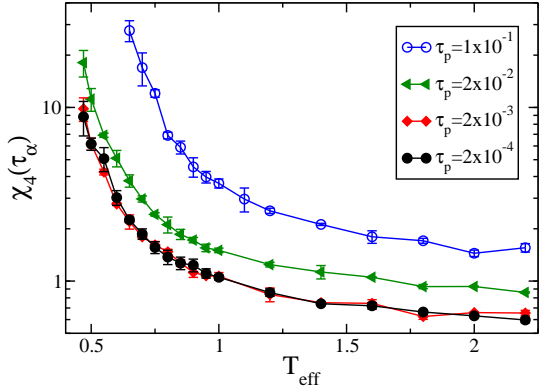}
\caption{\label{chiT} The increasing 
value of the four-point susceptibility $\chi_4(\tau_\alpha)$ 
at the $\alpha$-relaxation time $\tau_\alpha$ for different 
$\tau_p$ mirrors the increase of the relaxation time 
in Fig.~\ref{tau}.}
\end{figure}

Shown in Fig.~\ref{chiT} is the value of the dynamic susceptibility at 
the $\alpha$-relaxation time, 
$\chi_4(\tau_\alpha) \equiv  \chi_4(k;t=\tau_\alpha)$, 
as a function of the effective temperature, 
$T_{\mathrm{eff}}$, for the same $\tau_p$ values. 
The effective temperature dependence of 
$\chi_4(\tau_\alpha)$ shows similar trend as $\tau_\alpha$. 
There is a modest change for $T_{\mathrm{eff}} \ge 1.2$, but there is a sudden 
increase in $\chi_4(\tau_\alpha)$ for $T_{\mathrm{eff}}$, where we observe large 
increases in $\tau_\alpha$ at a fixed effective temperature. 
This simply reveals that slow dynamics is accompanied by 
the growth of the extent of 
spatially heterogeneous dynamics in self-propelled particle systems.

\begin{figure}
\includegraphics[width=3.2in]{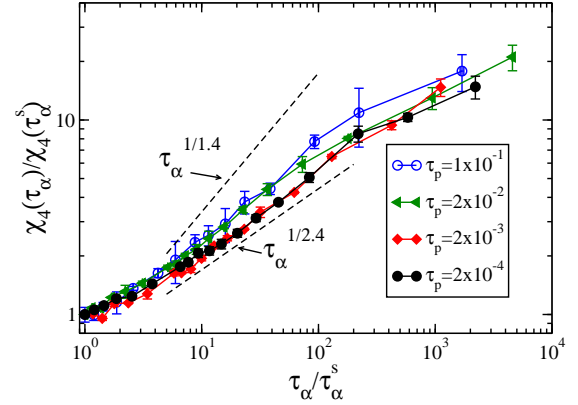}
\caption{\label{tauc4}
Evolution of the four-point susceptibility $\chi_4(\tau_\alpha)$
with $\tau_\alpha$ for various persistence times. Both axis are rescaled 
by their values at the high effective temperature 
value $T_{\rm eff}=2.2$. The remaining dependence on $\tau_p$ is qualitatively
accounted by the different values of the exponent $\gamma$
describing the intermediate mode-coupling regime.}
\end{figure}

To examine if the increase in $\chi_4(\tau_\alpha)$ is solely a 
reflection of the increase in $\tau_\alpha$, we replot 
$\chi_4(\tau_\alpha)$ versus $\tau_\alpha$ on a log-log scale in 
Fig.~\ref{tauc4}. To build this plot, we also rescaled 
$\chi_4(\tau_\alpha)$ and $\tau_\alpha$ by their high-temperature values,
denoted $\chi_4(\tau_\alpha^S)$ and $\tau_\alpha^S$, which 
we measure for convenience at the highest studied 
effective temperature, $T_{\rm eff}=2.2$.
This rescaling is a simple way to scale out once again 
the high-temperature behavior in the liquid, in order 
to focus our attention on the slowing down and the increase of dynamic
heterogeneity resulting from the glassy dynamics itself.  
Whereas the $\tau_p$-dependence of $\tau_\alpha^S$ was 
already discussed above, 
we attribute the increase of $\chi_4$ with $\tau_p$ in the liquid 
to the related increase of the spatial extent of the velocity 
correlations that was shown in Fig.~\ref{omega}(b). This observation
suggests that already in the high effective temperature active liquid
the motion is spatially correlated to some extent. Preliminary results suggest that
the spatial extent of the high effective temperature dynamic correlations 
increases with increasing persistence time but a more detailed investigation of
this phenomenon is left for a future study.

After rescaling of the high-temperature physics, the 
dependence of $\chi_4(\tau_\alpha)$ on $\tau_\alpha$ changes extremely 
weakly with the persistence time $\tau_p$, see Fig.~\ref{tauc4}. 
This observation is consistent with the idea that 
the glassy dynamics in and out of equilibrium are qualitatively 
very similar. 

In detail, $\chi_4(\tau_\alpha)$ first increases rather rapidly with increasing 
$\tau_\alpha$ in the initial regime of slowing down corresponding 
to the fitted mode-coupling regime in Fig.~\ref{mctfit}. For even 
larger relaxation times, the four-point susceptibility 
reaches a regime where the growth with $\tau_\alpha$ is 
much less pronounced, as observed in thermal 
systems \cite{Berthierbook,Dalle2007}.  
In the mode-coupling regime, one expects \cite{Toninelli2005} $\chi_4 \sim 
(T_{\rm eff}-T_{\rm eff}^c)^{-1}$. By combining this 
behavior to the power law description of the 
relaxation time, one then expects that $\chi_4 \sim \tau_\alpha^{1/\gamma}$
in the regime approximately described by mode-coupling theory inspired
fits. Since the exponent $\gamma$ in Fig.~\ref{mctparam}(b) decreases
from $\gamma=2.4$ to $\gamma=1.4$ when increasing $\tau_p$, we also 
add these two extreme cases as guides to the eye in Fig.~\ref{tauc4}.
While not terribly convincing, these power laws appear to account 
qualitatively correctly for the weak dependence on $\tau_p$ of the 
various series of data shown in Fig.~\ref{tauc4}.

We finish this section by noting that we examined mobility correlations
by studying a dynamic susceptibility that in thermal systems 
depends explicitly on the 
simulation ensemble \cite{Berthierbook,Berthier2005}. 
It should also be expected that the four-point 
susceptibility studied in this work would be different if we allowed for 
fluctuations of conserved quantities, for example 
if we allowed for volume 
fluctuations \cite{Flenner2011,Flenner2013,Flenner2014}. However, unlike 
for an equilibrium system, it is unclear how to account for fluctuations 
of conserved quantities in the nonequilibrium active systems. 
Studies of four-point structure factors for 
large systems may provide insight into the correlated dynamics of 
dense self-propelled systems since the large system size
would allow one to obtain the full dynamic susceptibility by extrapolating 
the small $q$ values of the four-point structure factor \cite{Flenner2011}. 
This investigation is beyond the scope of the present work.
 
\section{Summary and conclusions}

We studied the glassy dynamics of the Kob-Andersen binary
mixture of self-propelled particles. The
self-propulsion force follows the Ornstein-Uhlenbeck stochastic process, and 
there is no thermal noise.  Due to the lack of thermal noise, 
the dynamics can be characterized by a persistence time 
of the random force $\tau_p$, and an effective temperature
$T_{\mathrm{eff}}$ that is related to the long time diffusion coefficient
of an isolated particle. 

We found that the local structuring of the active fluid,
as seen via the pair correlation functions,
is significantly enhanced as one increases the
persistence time $\tau_p$. The enhancement can be examined through 
the height of the first peak of the pair correlation function of the larger,
more abundant $A$ particles, $g_{AA}(r)$. This peak height for 
the largest persistence time studied in this work, $\tau_p = 1\times10^{-1}$, 
is larger at $T_{\mathrm{eff}} = 2.0$, in the fluid phase, than for 
the smallest persistence time $\tau_p = 2\times10^{-4}$ 
at $T_{\mathrm{eff}} = 0.47$, in the glassy-dynamics regime. Furthermore, new 
features in the pair correlation functions, \textit{e.g.} the splitting of 
the second peak of $g_{AA}(r)$ into three sub-peaks, are
observed in the self-propelled system with large persistence times. 
These features suggests that the self-propulsion can favor structures 
that are not present in the equilibrium system. Therefore, the local 
structure of the self-propelled particles departs 
very strongly from that of particles in thermal equilibrium. 

More pronounced correlations of self-propelled particles' positions revealed by the 
pair distribution function are accompanied by correlations of the velocities
of the particles. These correlations are absent in equilibrium; 
they increase significantly
with increasing persistence time. The velocity correlations influence
the short-time dynamics of the active system. Within our mode-coupling-like theory \cite{Szamel2015,Szamel2016}, they also influence the long-time 
glassy dynamics. 

Importantly, we do not see a significant change in the structure 
with decreasing the effective temperature at a fixed persistence time.  
Therefore, it makes sense to 
study the glassy dynamics of systems with fixed persistence 
time and decreasing effective temperature. 
When the persistence 
time goes to zero the system behaves as an overdamped Brownian system
at a temperature equal to the effective temperature. We find that for 
a small increase of the persistence time, the dynamics initially
speeds up and then it slows down for larger persistence times at every effective
temperature. However, if we 
fit the high effective temperature results to an Arrhenius law
and examine the slowdown as deviations from this high temperature 
Arrhenius behavior, we find that the slowdown of the dynamics increases 
monotonically as the persistence time increases. Additionally, we find that
fits to a Vogel-Fulcher form of the relaxation time versus the
effective temperature also lead to a monotonic increase of both the 
glass transition temperature and the fragility parameter with increasing persistence time.
Fits to a mode-coupling theory power law also result in 
a monotonic increase of the mode-coupling transition temperature.

Overall, these findings suggest that once the 
(non-trivial and far-from-equilibrium) physics of the 
high-temperature fluid is scaled out, the resulting glassy dynamics 
evolves rather smoothly with an increase of the persistence time 
and does not show any striking qualitative feature that is not also
typically observed in standard models for equilibrium supercooled liquids. 
A remarkable conclusion is that for the present model, 
the increase of the persistence promotes glassy behavior, rather 
than suppresses it. These results therefore show that 
the idea \cite{Mandal2014,nandi}
that self-propulsion can be universally interpreted as a driving force 
acting against, and potentially suppressing, the equilibrium glassy 
dynamics is not correct.
We conclude instead that the interplay between self-propulsion and
glassy dynamics is much more subtle \cite{Berthier2013,Szamel2015,Szamel2016}
and the outcome cannot easily
be predicted on phenomenological grounds. Whereas for hard particles, 
the glass transition is depressed from its equilibrium 
limit \cite{Berthier2014,Ni2013}, 
the opposite behavior is found for the present Lennard-Jones model.
Capturing these features in an accurate microscopic theory 
represents therefore an important challenge for future work.

We also found that the Stokes-Einstein relation $D \sim \tau_\alpha^{-1}$ was
valid for small persistence times at effective temperatures above the onset
of supercooling. However, for the largest persistence times studied in 
this work, 
the Stokes-Einstein relation is never valid. At higher effective 
temperatures $D\tau_\alpha$ 
decreases, reaches a minimum, then begins to increase for effective 
temperatures when 
the liquid is supercooled. However, in the glassy regime 
corresponding to large relaxation times it appears that 
$D\tau_\alpha$ follows a master-curve, which provides 
additional support to the idea that the glassy dynamics of the 
supercooled self-propelled liquid shares similar characteristics to the 
dynamics of equilibrium supercooled liquids. 

We finished the study of glassy dynamics by examining correlations
in the relaxation of the individual particles, \textit{i.e.} dynamic heterogeneity. To this 
end we analyzed a dynamic susceptibility $\chi_4(k;t)$ which is a measure 
of the fluctuations of the self-intermediate scattering function. Since
our system is at constant density and concentration, our 
dynamic susceptibility should be considered a lower bound to the 
dynamic susceptibility that allows for all the fluctuations in the 
system \cite{Berthier2005,Flenner2011,Flenner2014,Flenner2013}. 
It is unclear how other fluctuations would contribute to
the dynamic susceptibility and future work should analyze this issue
in more detail. One possibility is to employ large-scale simulations 
to measure the low wave-vector behavior of a four-point 
dynamic structure factor. Again, the present work leads us to conclude that
when the high temperature regime is scaled 
out, the relation between $\chi_4(\tau_\alpha)$ and 
$\tau_\alpha$ appears rather insensitive to the persistence time. 

In conclusion, the detailed numerical
results presented here for the nonequilibrium 
glassy dynamics of self-propelled particles 
suggest that the structure and dynamics
of a `simple' (i.e. non-glassy)
dense active fluid change profoundly
with increasing departure from thermal
equilibrium, quantified by the persistence time of the self-propulsion. 
With
increasing $\tau_p$ correlations of the particles' positions become more 
pronounced and
spatial correlations of the velocities develop. 
The dynamics initially speed up 
but at longer
persistence time they slow down and dynamic correlations develop 
mirroring the correlations
of equal time velocities. These changes in the normal fluid structure and 
dynamics become more pronounced 
as the effective temperature decreases and the glassy regime is entered. 
However, while the details
of the slowing down depend quantitatively 
on the persistence time, the overall picture of glassy dynamics 
is largely similar 
for different persistence times. In other words, the evolution of the 
glassy dynamics 
with decreasing effective temperature in active systems is qualitatively 
similar to the
evolution of the dynamics in thermal glass-forming systems with decreasing 
the temperature. 

This broad conclusion implies that self-propelled particles 
undergo a glass transition at low effective temperatures 
or large densities accompanied by a complex time 
dependence of time correlation functions, 
locally-caged particle dynamics, and spatially heterogeneous dynamics
very much as in thermal equilibrium. Despite the local injection of 
energy \cite{Szamel2014,Fodor2016}
and violations of equilibrium fluctuation-dissipation 
relations \cite{Levis2015} the overall phenomenon 
studied here is therefore best described as a 
nonequilibrium glass transition~\cite{Berthier2013}.
While this conclusion is broadly consistent with the experimental 
reports of glassy dynamics in active materials, it remains 
to be understood whether the present model of self-propelled 
particles and the generic concept of a nonequilibrium glass transition
are sufficient to account for experimental observations. We hope that
future experiments using for instance self-propelled colloidal particles will 
clarify this issue.  

\section*{Acknowledgments} 
The research in Montpellier was supported by funding
from the European Research Council under the European
Union's Seventh Framework Programme (FP7/2007-2013) / ERC 
Grant agreement No 306845.
E. F. and G. S. gratefully acknowledge the 
support of NSF Grant No.~CHE 121340.


\begin{thebibliography}{99}

\bibitem{Fily2012}
Y. Fily, and M. C. Marchetti,
Phys. Rev. Lett. \textbf{108}, 235702 (2012). 

\bibitem{Cates2013}
M. E. Cates, and J. Tailleur,
Europhys. Lett. \textbf{101}, 20010. 

\bibitem{Redner2013a}
G. S. Redner, M. F. Hagan, and A. Baskaran,
Phys. Rev. Lett. \textbf{110}, 055701 (2013).

\bibitem{Redner2013}
G. S. Redner, A. Baskaran, and M. F. Hagan,
Phys. Rev. E \textbf{88}, 012305 (2013). 

\bibitem{Fily2014}
Y. Fily, S. Henkes, and M. C. Marchetti,
Soft Matter, \textbf{10}, 2132 (2014). 

\bibitem{Wysocki2014}
A. Wysocki, R. G. Winkler, and G. Gompper,
Europhys. Lett. \textbf{105}, 48004 (2014). 

\bibitem{Tailleur2008}
J. Tailleur and M. E. Cates, Phys. Rev. Lett. \textbf{100}, 218103 (2008).

\bibitem{Speck2014}
T. Speck, J. Bialk\'{e}, A. M. Menzel, and H. L\"owen, Phys.
Rev. Lett. \textbf{112}, 218304 (2014).

\bibitem{Takatori2014} 
S. Takatori, W. Yan, and J. Brady, Phys. Rev. Lett. \textbf{113},
028103 (2014).

\bibitem{Wittkowski2014}
R. Wittkowski, A. Tiribocchi, J. Stenhammar, R. J. Allen,
D. Marenduzzo, and M. E. Cates, Nature communications 5
(2014).

\bibitem{Solon2015}
A. P. Solon, J. Stenhammar, R. Wittkowski, M. Kardar, Y. Kafri,
M. E. Cates, and J. Tailleur, Phys. Rev. Lett. 114, 198301
(2015).

\bibitem{Farage2015}
T. F. F. Farage, P. Krinninger, and J. M. Brader, Phys. Rev. E
\textbf{91}, 042310 (2015).

\bibitem{Bialke2015}
J. Bialk\'{e}, J. T. Siebert, H. L\"owen, and T. Speck, Phys. Rev.
Lett. \textbf{115}, 098301 (2015).

\bibitem{Fodor2016}
E. Fodor, C. Nardini, M. E. Cates, J. Tailleur, P. Visco, and F. van Wijland,
arXiv:1604.00953.

\bibitem{Angelini2011}
T. E. Angelini, E. Hannezo, X. Trepat, M. Marques, J. J. Fredberg, 
and D. A. Weitz,
Proc. Natl. Acad. Sci. \textbf{108}, 4714 (2011). 

\bibitem{Schotz2013}
E.-M. Sch\"otz, M. Lanio, J. A. Talbot, and M. L. Manning,
J. R. Soc. Interface \textbf{10}, 20130726 (2013). 

\bibitem{Garcia2015}
S. Garcia, E. Hannezo, J. Elgeti, J.-F Joanny, P. Silberzan, and N. S. Gov,
Proc. Natl. Acad. Sci. \textbf{112}, 15314 (2015). 

\bibitem{Gravish2015}
N. Gravish, G. Gold, A. Zangwill, M. A. D. Goodisman, and D. I. Goldman,
Soft Matter \textbf{11}, 6552 (2015). 

\bibitem{alberto}
M. Tennenbaum, Z. Liu, D. Hu, A. Fernandez-Nieves, 
Nature Materials {\bf 15}, 54 (2016).

\bibitem{Henkes2011}
S. Henkes, Y. Fily, and M. C. Marchetti,
Phys. Rev. E \textbf{84}, 040301 (2011). 

\bibitem{Berthier2013}
L. Berthier and J. Kurchan,
Nature Phys. \textbf{9}, 310 (2013). 

\bibitem{Ni2013}
R. Ni, M. A. Cohen Stuart, and M. Dijkstra,
Nature Comm. \textbf{4}, 2704 (2013). 

\bibitem{Berthier2014}
L. Berthier,
Phys. Rev. Lett. \textbf{112}, 220602 (2014). 

\bibitem{Levis2015}
D. Levis and L. Berthier,
EPL {\bf 111}, 60006 (2015).

\bibitem{Szamel2015}
G. Szamel, E. Flenner, and L. Berthier,
Phys. Rev. E \textbf{91}, 062304 (2015). 

\bibitem{Szamel2016}
G. Szamel, Phys. Rev. E \textbf{93}, 012603 (2016).

\bibitem{chinese}
H. Ding, M. Feng, H. Jiang, Z. Hou,
arXiv:1506.02754.

\bibitem{Mandal2014}
R. Mandal, P. J. Bhuyan, M. Rao, and C. Dasgupta, arXiv:1412.1631

\bibitem{Eaves2014}
K. R. Pilkiewicz and J. D. Eaves,
Soft Matter {\bf 10}, 7495 (2014).

\bibitem{Reichhardt2014}
C. Reichhardt and C. J. Olson Reichhardt,
Phys. Rev. E {\bf 90}, 012701 (2014).

\bibitem{Manning2015}
D. Bi, J. H. Lopez, J. M. Schwarz, and M. L. Manning,
Nature Phys. {\bf 11}, 1074 (2015).

\bibitem{farage}
T. F. F. Farage and J. M. Brader,
arXiv:1403.0928.

\bibitem{nandi}
S. K. Nandi, arXiv:1605.06073.

\bibitem{Kob1994}
W. Kob and H. C. Andersen, Phys. Rev. Lett. \textbf{73}, 1376 (1994).

\bibitem{Szamel2014} 
G. Szamel, Phys. Rev E \textbf{90}, 012111 (2014).

\bibitem{Maggi2015} 
C. Maggi, U. M. B. Marconi, N. Gnan, and R. Di Leonardo,
Sci. Rep. \textbf{5}, 10742 (2015).

\bibitem{tenHagen}
B. ten Hagen, S. van Teeffelen, and H. L\"owen, 
J. Phys.: Condens. Matter \textbf{23}, 194119 (2011). 

\bibitem{runandtumble}
J. Tailleur and M. E. Cates,
Phys. Rev. Lett. {\bf 100}, 218103 (2008). 

\bibitem{Marconi2016}
U. M. B. Marconi, N. Gnan, M. Paoluzzi, C. Maggi, and R. Di Leonardo,
Scientific Reports \textbf{6}, 23297 (2016). 

\bibitem{Flenner2005a}
E. Flenner and G. Szamel,
Phys. Rev. E \textbf{72}, 011205 (2005). 

\bibitem{Flenner2005b}
E. Flenner and G. Szamel,
Phys. Rev. E \textbf{72}, 031508 (2005).

\bibitem{simple}
J.-P. Hansen and I. R. McDonald, \textit{Theory of Simple Liquids},
(Elsevier, 2012).

\bibitem{paddyreview}
C. P. Royall and S. R. Williams,
Phys. Rep. {\bf 560}, 1 (2015).

\bibitem{Berthier2010} See, however,
L. Berthier and G. Tarjus, Phys. Rev. E \textbf{82}, 031502 (2010), 
where it was argued
that the glassy dynamics may be sensitive to structural changes 
not reflected in the 
pair distribution function.

\bibitem{Goetzebook} W. G\"otze,
\textit{Complex dynamics of glass-forming liquids: A mode-coupling theory}
(Oxford University Press, Oxford, 2008).

\bibitem{Berthier2009}
L. Berthier and G. Tarjus, Phys. Rev. Lett. \textbf{103}, 170601 (2009).

\bibitem{Berthier2011}
L. Berthier and G. Biroli,
Rev. Mod. Phys. \textbf{83}, 587 (2011).

\bibitem{Berthierbook}
\textit{Dynamical Heterogeneities in Glasses, Colloids, and Granular Media},
edited by L. Berthier, G. Biroli, J.-P. Bouchaud, 
L. Cipelletti, and W. van Saarloos
(Oxford University Press, New York, 2011). 

\bibitem{Ediger2000}
M. D. Ediger,
Annu, Rev. Phys. Chem \textbf{51}, 99 (2000).

\bibitem{Flenner2011}
E. Flenner, M. Zhang, and G. Szamel,
Phys. Rev. E \textbf{83}, 051501 (2011).

\bibitem{Dalle2007}
C. Dalle-Ferrier, C. Thibierge, C. Alba-Simionesco, L. Berthier, 
G. Biroli, J.-P. Bouchaud, F. Ladieu, D. L'Hote, and G. Tarjus
Phys. Rev. E {\bf 76}, 041510 (2007).

\bibitem{Toninelli2005}
C. Toninelli, M. Wyart, L. Berthier, G. Biroli, and J.-P. Bouchaud,
Phys. Rev. E {\bf 71}, 041505 (2005).

\bibitem{Berthier2005}
L. Berthier, G. Biroli, J.-P. Bouchard, L. Cipelletti, 
D. El Masri, D. L'H\^ote, F. Ladieu, and M. Pierno,
Science \textbf{310}, 1797 (2005). 

\bibitem{Flenner2013}
E. Flenner and G. Szamel,
J. Chem. Phys. \textbf{138}, 12A523 (2013). 

\bibitem{Flenner2014}
E. Flenner, H. Staley, and G. Szamel,
Phys. Rev. Lett. \textbf{112}, 097801 (2014). 

\end{thebibliography}
\end{document}